\documentstyle[prl,aps,twocolumn,epsf]{revtex}
\newcommand*{\be}{\begin{equation}} 
\newcommand*{\ee}{\end{equation}}

\begin{document}

\title{ \bf  Storage of energy in confined quantum systems}
 
\author{A.P.C. Malbouisson${}^{(a)}$} 

\address{ (a) {\it CBPF/MCT - Rua Dr. Xavier Sigaud 150, Urca,} \\   
{\it Rio de Janeiro CEP 22290-180-RJ, Brazil. E-mail: adolfo@cbpf.br}}

\vspace{0.7cm}

\maketitle

\begin{abstract}

Using the non-perturbative method of {\it dressed} states  introduced in  previous publications, 
 we study the evolution  of a confined quantum mechanical system embedded in a {\it ohmic} environment.
  Our 
approach furnishes a theoretical mechanism to control inhibition of the decay of excited quantum  systems in cavities,  in both weak and strong coupling regimes. PACS number(s): ~03.65.Ca, 32.80.Pj 

\vspace{0.34cm}
\noindent

\end{abstract}

In recent publications \cite{adolfo1,adolfo2} a method  ({\it dressed} states) has been introduced that allows a  non-perturbative
approach to  situations where a material body (an atom or a Brownian particle) is coupled to a field, 
 provided 
that the interaction between the parts of the system can be approximated by
a linear coupling. From now on we use the term "particle" in a general sense, 
refering to an atom, a Brownian particle a molecule, etc.... Our {\it dressed} 
states can be viewed as a rigorous version of the semiqualitative idea of dressed atom 
introduced originally in ref.\cite{Haroche}, which can be constructed in reason of the linear character of our problem. 
More precisely, the method applies for all systems that can be 
described by an Hamiltonian of the form,
\begin{equation} 
H=\frac{1}{2}\left[p_{0}^{2}+\omega_{0}^{2}q_{0}^{2}+ 
\sum_{k=1}^{N}(p_{k}^{2}+\omega_{k}^{2}q_{k}^{2}\right]-q_{0}\sum_{k=1}^{N}c_{k}q_{k},  
\label{Hamiltoniana} 
\end{equation} 
where the subscript $0$ refers to the "material body" and $k=1,2,...N$ refer to the harmonic environment modes.  
A Hamiltonian of this type, describing a linear coupling 
 of a particle with an environment, has been used in \cite {paz} to study 
the quantum Brownian motion of a particle with the path-integral formalism.  
  The limit $N\rightarrow \infty$ in Eq.(\ref{Hamiltoniana}) is understood. We investigate also the behaviour of the system as a 
function of the strenght of the coupling between the particle and the bath. In particular
we give explicitly non-perturbative formulas for the decay probability of the particle in the weak and strong coupling regimes.

We consider the problem of a particle $q_{0}$ in the harmonic approximation 
 having (bare) frequency $\omega_{0}$ 
coupled to $N$ other oscillators $q_{i}$ of frequencies $\omega_{i},\;\;i=1,2,...N$.
 In the limit $N\rightarrow \infty$ we recover 
our situation of the coupling particle-bath (field) after redefinition of divergent
quantities.  The 
bilinear Hamiltonian (\ref{Hamiltoniana}) can be turned to principal axis by means of a point 
transformation, 
$q_{\mu}=t_{\mu}^{r}Q_{r}\;\;\;\;\;p_{\mu}=t_{\mu}^{r}P_{r};\;\; \mu=(0,\{k\}),\;k=1,2,..., N;\;\; 
r=0,...N,$ 
performed 
by an orthonormal  matrix $T=(t_{\mu}^{r})$. The subscript $\mu=0$ and $\mu=k$ refer respectively to the particle and the 
harmonic modes  of the bath and $r$ refers to the normal modes. In terms of 
normal momenta and coordinates, the transformed 
Hamiltonian in principal axis reads,
$H=\frac{1}{2}\sum_{r=0}^{N}(P_{r}^{2}+\Omega_{r}^{2}Q_{r}^{2}),$
 where the $\Omega_{r}$'s are the normal frequencies corresponding to the possible 
collective stable oscillation modes of the coupled system. The matrix elements $t_{\mu}^{r}$ 
are given by \cite{adolfo1}
\begin{equation} 
t_{k}^{r}=\frac{c_{k}}{(\omega_{k}^{2}-\Omega_{r}^{2})}t_{0}^{r}\;,\;\; 
t_{0}^{r}= \left[1+\sum_{k=1}^{N}\frac{c_{k}^{2}}{ 
(\omega_{k}^{2}-\Omega_{r}^{2})^{2}}\right]^{-\frac{1}{2}}
\label{tkrg1} 
\end{equation} 
with the condition,
\begin{equation} 
\omega_{0}^{2}-\Omega_{r}^{2}=\sum_{k=1}^{N}\frac{c_{k}^{2}}
{\omega_{k}^{2} -\Omega_{r}^{2}}.  
\label{Nelson1} 
\end{equation}

We take $c_{k}=\eta (\omega_{k})^{n}$. In this case the environment is classified according to 
$n>1$, $n=1$, or $n<1$, respectively as {\it supraohmic}, {\it ohmic} or {\it subohmic}. 
For a subohmic environment the sum in Eq.(\ref{Nelson1}) is convergent and the frequency 
$\omega_{0}$ is well defined. For ohmic and supraohmic environments the sum 
in the right hand side of 
Eq.(\ref{Nelson1}) diverges what makes the equation meaningless as it stands,
 a renormalization 
procedure being needed. 
 In this case, by a straightforward generalization of the method described in \cite{adolfo1} we can define a {\it renormalized} 
frequency $\bar{\omega}$, by means of a counterterm $\delta \omega^{2}$,
\be
\bar{\omega}^{2}=\omega_{0}^{2}-\delta \omega^{2}\;\;;\delta \omega^{2}=\frac{\eta^{2}}{4}\sum_{k=1}^{N}\sum_{\alpha=1}^{E[n]}\Omega_{r}^{2(\alpha-1)}\omega_{k}^{2(n-\alpha)},
\label{omegabarra}
\ee
in terms of which Eq.(\ref{Nelson1}) becomes,
\be
 \bar{\omega}^{2}-\Omega_{r}^{2}=\eta^{2}\sum_{k=1}^{N}\frac{\Omega_{r}^{2E[n]}}
{\omega_{k}^{2} -\Omega_{r}^{2}},  
\label{Nelson3} 
\ee
with the notation $E[n]$ standing for the smallest integer containing $n$.
We see that in the limit $N\rightarrow \infty$ the above procedure is exactly the analogous 
of naive mass renormalization in Quantum Field Theory: the addition of a counterterm 
$-\delta \omega^{2}q_{0}^{2}$ allows to compensate the infinuty of $\omega_{0}^{2}$ in such 
a way as to leave a finite, physically meaninful renormalized frequency $\bar{\omega}$. 
This simple renormalization scheme has been originally introduced in ref.\cite{Thirring}.

To proceed, we take the constant $\eta$ as   
$\eta=\sqrt{2g\Delta\omega}$, $\Delta\omega$ being the interval between two neighbouring
bath frequencies (supposed uniform) and where $g$ is some constant [with dimension of 
$(frequency)^{2-\eta}$].
For reasons that will become apparent later, we restrict ouselves to 
 the physical situations in which the whole system is 
confined to a cavity of diameter $L$, and the environment frequencies $\omega_k$ 
can be writen in the form 
\be
\omega_k=2k\pi/L,\;\;\;\;k=1,2,...\;.
\label{discreto}
\ee
 Then 
 using the formula, 
\begin{equation}
\sum_{k=1}^{N}\frac{1}{(k^{2}-u^{2})}= \left[\frac{1}{2u^{2}}-\frac{\pi}{u}
{\rm cot}(\pi u)\right],
\label{id4}
\end{equation}
and restricting ourselves to an {\it ohmic} environment, 
Eq.(\ref{Nelson3}) can be written in closed form,
\begin{equation} 
\mathrm{cot}(\frac{L\Omega}{2c})=\frac{\Omega}{\pi g}+\frac{c}
{L\Omega}(1-\frac{\bar{\omega}^{2}L}{\pi gc}).  
\label{eigenfrequencies1} 
\end{equation}
The solutions of Eq.(\ref{eigenfrequencies1})  
with respect to  $\Omega$ give the 
spectrum of eigenfrequencies $\Omega_{r}$ corresponding to the collective normal modes.\\
The transformation matrix elements turning the material body-bath system to principal 
axis is obtained in terms of the physically meaningful quantities $\Omega_{r}$, $\bar{\omega}$, 
 after some rather long but straightforward manipulations analogous as 
it has been done in \cite{adolfo1}. They read,  
\begin{eqnarray} 
t_{0}^{r}&=&\frac{\eta \Omega_{r}}{\sqrt{(\Omega_{r}^{2}-\bar{ 
\omega}^{2})^{2}+\frac{\eta^{2}}{2}(3\Omega_{r}^{2}-\bar{\omega}^{2})+
\pi^{2}g^{2}\Omega_{r}^{2}}}\;, \nonumber \\   
& & t_{k}^{r}=\frac{\eta\omega_{k}}{\omega_{k}^{2}-\Omega_{r}^{2}}t_{0}^{r}. 
\label{t0r2} 
\end{eqnarray}

To study the time evolution of the particle,  
we start from the eigenstates of our system, represented by the normalized 
eigenfunctions,  
\begin{eqnarray} 
\phi_{n_{0}n_{1}n_{2}...}(Q,t)&=&\prod_{s}\left[\sqrt{\frac{2^{n_s}}{n_s!}}H_{n_{s}}(\sqrt{\frac{ 
\Omega_{s}}{\hbar}}Q_{s})\right]\times\nonumber\\
& &~~~~~~\Gamma_{0}e^{-i\sum_{s}n_{s}\Omega_{s}t}, 
\label{autofuncoes} 
\end{eqnarray} 
where $H_{n_{s}}$ stands for the $n_{s}$-th Hermite polynomial 
and $\Gamma_{0}$ is the normalized vacuum eigenfunction.  
We introduce {\it dressed} coordinates $q^{\prime}_{0}$ and $ 
\{q^{\prime}_{i}\}$ for, respectively the {\it dressed} atom and  
the {\it dressed} field, defined by \cite{adolfo1},  
\begin{equation} 
\sqrt{\frac{\bar{\omega}_{\mu}}{\hbar}}q^{\prime}_{\mu}=\sum_{r}t_{\mu}^{r} 
\sqrt{\frac{\Omega_{r}}{\hbar}}Q_{r},  
\label{qvestidas1} 
\end{equation} 
valid for arbitrary $L$ and where $\bar{\omega}_{\mu}=\{\bar{\omega}, 
\;\omega_{i}\}$,  and let us  define for a fixed instant the complete orthonormal set of functions \cite{adolfo1},  
\begin{equation} 
\psi_{\kappa_{0} \kappa_{1}...}(q^{\prime})=\prod_{\mu}\left[\sqrt{\frac{2^{\kappa_{\mu}}}{\kappa_{\mu}!}} 
H_{\kappa_{\mu}} (\sqrt{\frac{\bar{\omega}_{\mu}}{\hbar}} 
q^{\prime}_{\mu})\right]\Gamma_{0},  
\label{ortovestidas1} 
\end{equation} 
where $q^{\prime}_{\mu}=q^{\prime}_{0},\, q^{\prime}_{i}$, $\bar{\omega} 
_{\mu}=\{\bar{\omega},\, \omega_{i}\}$. Note that 
the ground state $\Gamma_{0}$ in the above equation 
is the same as in Eq.(\ref{autofuncoes}). The invariance of the ground state 
 is due to our definition of {\it dressed} coordinates given by Eq.
(\ref{qvestidas1}). Each function $\psi_{\kappa_{0} \kappa_{1}...}(q^{\prime})$ describes a state in which the {\it dressed} oscillator 
$q'_{\mu}$ is in its $\kappa_{\mu}-th$ excited state.
Using Eq.(\ref{qvestidas1}) the functions  
(\ref{ortovestidas1}) can be expressed in terms of the normal coordinates  
$Q_{r}$. But since (\ref{autofuncoes}) is a complete set of orthonormal 
functions, the functions (\ref{ortovestidas1}) may be written as linear 
combinations of the eigenfunctions of the coupled system, and conversely.  
 
 Let us call $\Gamma_{1}^{0}=\psi_{100...0(\mu)0...}$ the configuration 
in which only the 
dressed oscillator $q^{\prime}_{0}$ (the {\it dressed} particle) is in the first excited level. Then it is shown in 
\cite{adolfo1}  
 the following expression for the time evolution of the first-level excited dressed 
particle $q^{\prime}_{0}$,  
\begin{equation} 
\Gamma_{1}^{0}(t)=\sum_{\nu}f^{0 \nu}(t)\Gamma_{1}^{\nu}(0)\;;\;\;\;\; f^{0 \nu}(t)=\sum_{s}t_{0}^{s}t_{\nu}^{s}e^{-i\Omega_{s}t}.
\label{ortovestidas5} 
\end{equation} 
From Eq.(\ref{ortovestidas5}) we see that 
 the initially excited dressed particle naturally distributes its 
energy among itself and all other dressed oscillators (the environment) as time goes on, 
with probability amplitudes given by the quantities 
 $f_{0 \nu}(t)$ in Eq.(\ref{ortovestidas5}). In Eq.(\ref{ortovestidas5}) the coefficients $f^{0 \nu}(t)$ have a simple 
interpretation: $f^{00}(t)$ 
and $f^{0i}(t)$  are respectively the probability amplitudes that at time $t 
$ the dressed particle still be excited or have radiated a 
 quantum of frequency $h\omega_{i}$. We see that this formalism allows a 
quite natural description of the radiation process as a simple exact time 
evolution of the system. In the case of a very large cavity (free space) our method 
reproduces for weak coupling the well-known perturbative results \cite{adolfo1,adolfo2}. 
Nevertheless it should be remarked that due to our approximations to be done below for the confined system, we {\it can} {\it not} recover free space results from our confined system. 

Next we study the time evolution  process of the confined {\it dressed} particle 
when it is prepared in such a way that initially it is in its first excited state.
We shall consider the situation of  the particle 
confined  in a cavity of diameter $L$.

Let us now consider the {\it ohmic} system in which the particle is placed in the center
of a cavity of diameter $L$, in the case of  a small $L$, $i.e.$ that satisfies the 
condition of being much smaller than the coherence lenght, $L<<2c/g$. We note that from a 
physical point of view, $L$ stands for either the diameter of a spherical cavity or the 
spacing between infinite paralell mirrors. To obtain the eigenfrequencies spectrum, we remark that from a graphical analysis 
of Eq.(\ref{eigenfrequencies1}) it can be seen that in the case of small values of $L$, 
its  solutions are very near the frequency values corresponding to the asymptots of the
curve $\mathrm{cot}(\frac{L\Omega}{2c})$, which correspond to the environment modes 
$\omega_{i}=i2\pi c/L$, except from the smallest eigenfrequency $\Omega_{0}$. As we take larger and larger solutions,
they are nearer and nearer to the values corresponding to the asymptots. For instance, 
for a value of $L$ of the order of $2\times 10^{-2}m$  and $\bar{\omega}\sim 10^{10}/s$, 
only the lowest eigenfrequency $\Omega_{0}$ is significantly different from the field 
frequency corresponding to the first asymptot, all the other eigenfrequencies $\Omega_{k},
\;k=1,2,...$ being very close to the field modes $k2\pi c/L$. For higher values of 
$\bar{\omega}$ (and lower values of $L$) the differences between the eigenfrequencies and
the field modes frequencies are still smaller. Thus to solve Eq.(\ref{eigenfrequencies1})
for the larger eigenfrequencies we expand the function $\mathrm{cot}(\frac{L\Omega}{2c})$ 
around the values corresponding to the asymptots. We write,
\begin{equation}
\Omega_k=\frac{2\pi c}{L}(k+\epsilon_k),~~~k=1,2,..
\label{others}
\end{equation} 
with $0<\epsilon_{k}<1$, satisfying the equation,
\begin{equation}
\mathrm{cot}(\pi \epsilon_k)=\frac{2c}{gL}(k+\epsilon_k) +\frac{1}{(k+\epsilon_k)}
(1-\frac{\bar{\omega}^{2}L}{2\pi gc}).  
\label{eigen2} 
\end{equation}
But since for a small value of $L$ every $\epsilon_k$ is  much smaller than $1$, 
Eq.({\ref{eigen2}) can be linearized in  $\epsilon_k$, giving, 
\begin{equation}
\epsilon_k=\frac{4\pi g c L k}{2(4\pi^{2} c^{2} k^{2}-\bar{\omega}^{2}L^{2}}.
\label{linear}
\end{equation}
Eqs.(\ref{others}) and (\ref{linear}) give approximate solutions to the eigenfrequencies 
$\Omega_{k},\;\;k=1,2...$.

To solve Eq.(\ref{eigenfrequencies1}) with respect to the lowest eigenfrequency 
$\Omega_{0}$, let us assume that it satisfies the condition $\Omega_{0}L/2c<<1$ 
(we will see below that this condition is compatible with the condition of a small $L$ as
defined above). Inserting the condition $\Omega_{0}L/2c<<1$ in
Eq.(\ref{eigenfrequencies1}) and keeping up to quadratic terms in $\Omega$ we obtain the 
solution for the lowest eigenfrequency  
$\Omega_{0}$,
\begin{equation}
\Omega_0=\frac{\bar{\omega}}{\sqrt{1+\frac{\pi gL}{2c}}}.
\label{firsts}
\end{equation}
Consistency between Eq.(\ref{firsts}) and the condition  $\Omega_{0}L/2c<<1$ gives 
a condition on $L$,
\begin{equation}
L\ll 
\frac{2c}{g}f\;;\;\;\;f=\frac{\pi}{2}\left(\frac{g}{\bar{\omega}}\right)^2
\left(1+\sqrt{ 1+\frac{4}{\pi^2}\left(\frac{\bar{\omega}}{g}\right)^2}~\right).
\label{rsmall}
\end{equation}
Let us consider  
the situations of {\it weak} coupling ($g\ll \bar{\omega}$) and 
of {\it strong} coupling ($g\gg \bar{\omega}$), and let us consider the situation where the dressed material body is initially 
in its first excited level. Then from Eq.(\ref{ortovestidas5})  we obtain the probability that it 
will still be excited after a ellapsed time $t$, 
\begin{eqnarray}
|f^{00}(t)|^{2}&=&(t_{0}^{0})^{4}+2\sum_{k=1}^{\infty}(t_{0}^{0})^{2}
(t_{0}^{k})^{2}\cos(\Omega_{k}-\Omega_{0})t+\nonumber\\
& & ~~~~~~~\sum_{k,l=1}^{\infty}(t_{0}^{k})^{2}
(t_{0}^{l})^{2}\cos(\Omega_{k}-\Omega_{l})t.
\label{|f00R|2}
\end{eqnarray}

{\it a)} {\it Weak} {\it coupling}\\
In the case of {\it weak} coupling a physically interesting situation is when interactions
of electromagnetic type are involved. In this case, we take $g=\alpha \bar{\omega}$, where $\alpha$
is the fine structure constant, $\alpha=1/137$. Then the factor $f$ multiplying $2c/g$ in
Eq.(\ref{rsmall}) is $\sim 0.07$ and the condition $L\ll 2c/g$ is replaced by a more 
restrictive one, $L\ll 0.07(2c/g)$. For a typical infrared frequency, for instance 
$\bar{\omega}\sim 2,0\times 10^{11}/s$, our calculations are valid for a value of $L$,   
$L\ll 10^{-3}m$.\\

From Eqs.(\ref{t0r2}) and using the above expressions for the eigenfrequencies for small 
$L$, we obtain the matrix elements,
\begin{equation}
(t_0^0)^2\approx 1-\frac{\pi g L}{2c};\;\;(t_0^k)^2\approx \frac{g L}{\pi c k^2}.
\label{too}
\end{equation}
To obtain the above equations we have neglected the corrective term $\epsilon_{k}$, from 
the expressions for the eigenfrequencies $\Omega_{k}$. Nevertheless, corrections in 
$\epsilon_{k}$ should be included in the expressions for the matrix elements $t_{k}^{k}$,
in order to avoid spurious singularities due to our approximation.

Using Eqs.(\ref{too}) in Eq.({\ref{|f00R|2}), we obtain
 
\begin{eqnarray}
|f^{00}(t)|^{2}&\approx &1-\pi \delta+4(\frac{\delta}{\pi}-\delta^{2})
\sum_{k=1}^{\infty}\frac{1}{k^{2}}\cos(\Omega_{k}-\Omega_{0})t+\nonumber\\
& &~~~\pi^{2}\delta^{2}+\frac{4}{\pi^{2}}\delta^{2}\sum_{k,l=1}^{\infty}
\frac{1}{k^{2}l^{2}}\cos (\Omega_{k}-\Omega_{l})t,
\label{f002}
\end{eqnarray}
where we have introduced the dimensionless parameter $\delta=Lg/2c$, corresponding
to a small value of $L$ and we remember that the eigenfrequencies are given by 
Eqs.(\ref{others}), (\ref{linear}) and (\ref{firsts}). As time goes on, the probability that the 
material body be excited  attains periodically a minimum value which has a lower 
bound given by,

\begin{equation}
\mathrm{Min}(\delta)=1-\frac{5\pi}{3}\delta+\frac{14\pi^{2}}{9}\delta^{2}.
\label{min}
\end{equation}
For instance, for a frequency $\bar{\omega}$ of the order $\bar{\omega}\sim 4.00\times 10^{14}/s$ 
(in the red visible), and $L\sim 1.0\times 
10^{-6}m$, remembering our weak coupling {\it anzatz} $g=\bar{\omega}\alpha$, 
we see from Eq.(\ref{min}) that the probability that the material body be at 
any time excited will never fall below a value $\sim 0.97$, or a decay probability that
is never higher that a value $\sim 0.03$. It is interesting to compare this result with 
experimental observations in \cite{Jhe}, where stability is found for atoms 
emitting in the visible range placed between two parallel mirrors a distance 
$L=1.1\times 10^{-6}m$ apart from one another \cite{Jhe}.  For 
cavities small enough the probability that the material body remain excited as 
time goes on, oscillates with time between a maximum and a minimum values and never 
departs significantly from the situation of stability  in the excited state.

Nevertheless, in the weak coupling regime for any fixed value of the emission 
frequency,  
the probability that the particle remain 
excited can be considerably changed. This is a general phenomenon. We see from Eq. (\ref{min}) that mathematically there are {\it two} 
values of the cavity size, corresponding to $\delta =0$ and $\delta =15/14\pi$ which correspond 
to the probability {\it one} of the particle be excited at any time. Also, we can see from 
Eq.(\ref{min}) that the minimal value of $\mathrm{Min}(|f^{00}(t)|^{2})$ is attained for 
$\delta=15/28\pi$. The size of the cavity corresponding to this value of $\delta$ is exactly 
{\it half} the (non-zero) size of the cavity allowing from a mathematical point of view, permanent excitation of the particle. However we should not give physical meaning to the portion 
of the parabole (\ref{min}) from $\delta=15/28\pi$ to infinity, it will correspond for values 
$\delta >15/14\pi$ to probabilities larger than one and is physically meaningless. The physically meaningful part of the parabole (\ref{min}) corresponds to $0<\delta<15/28\pi$. Our conclusion is that 
 varying $\delta$ from $0$ to $\delta=15/28\pi$  we attain at the middle of the curve the point of maximum emission, where $\mathrm{Min}(|f^{00}(t)|^{2})\approx 0.55$. In the weak coupling regime, this is the minimal possible probability to the particle remain excited inside the cavity for any emmission frequency.

{\it b)} {\it Strong} {\it coupling}\\
In this case it can be seen from Eqs (\ref{linear}), (\ref{firsts}) and (\ref{rsmall}) that $\Omega_{0}\approx \bar{\omega}$ 
 and we 
obtain from Eq.(\ref{t0r2}),
\begin{equation}
(t_0^0)^2\approx \frac{1}{1+\pi \delta/2};\;\;(t_0^k)^2\approx \frac{g L}{\pi c k^2}.
\label{toos}
\end{equation}
Using Eqs.(\ref{toos}) in Eq.({\ref{|f00R|2}), we obtain for the probability that the 
excited system be still at the first energy level at time $t$, the expression,
\begin{eqnarray}
 |f^{00}(t)|^{2}&\approx &\left(\frac{2}{2+\pi \delta}\right)^{2}+\frac{2}
 {2+\pi \delta}\sum_{k=1}^{\infty}\frac{2\delta}{\pi k^{2}}
 \cos(\Omega_{k}-\Omega_{0})t+\nonumber\\
& &~~~+\frac{4}{\pi^{2}}\delta^{2}\sum_{k,l=1}^{\infty}\frac{1}{k^{2}l^{2}}
\cos (\Omega_{k}-\Omega_{l})t.
\label{f002s}
\end{eqnarray}
We see from (\ref{f002s}) that as the system evolves in time, the probability that the material body  
be excited, attains periodically a minimum value which has a lower bound given by,
\be
\mathrm{Min}(\delta)=\left(\frac{2}{2+\pi \delta}\right)^{2}-
\left(\frac{2}{2+\pi \delta}\right)\frac{\pi \delta}{3}-\frac{\pi^{2}\delta^{2}}{9}.
\label{mins}
\ee
The behaviour of the system in the strong coupling regime is completely different from 
weak coupling behaviour.
The condition of positivity of (\ref{mins}) imposes for {\it fixed} values of $g$ and 
$\bar{\omega}$ an upper bound for the quantity $\delta$, $\delta_{max}$, which corresponds
to an upper bound to the diameter $L$ of the cavity, $L_{max}$ (remember $\delta=Lg/2c$).
Values of $\delta$ larger than $\delta_{max}$, or equivalently, values of $L$ larger than 
$L_{max}$ are unphysical (correspond to negative probabilities) 
and should not be considered. These upper bounds are obtained 
from the solution with respect to $\delta$, of the inequality,
$\mathrm{Min}(\delta)\geq 0$.
We have $\mathrm{Min}(\delta)>0$ or $\mathrm{Min}(\delta)=0$, for 
respectively $\delta<\delta_{max}$ or $\delta =\delta_{max}$. The solution of the equation 
$\mathrm{Min}(\delta)=0$ gives,
$\delta_{max}=-\frac{1}{\pi}+\frac{\sqrt{-2+3\sqrt{5}}}{\pi}\approx 0.37.$
For  $\delta=\delta_{max}$, the minimal probability that the atom remain stable vanishes. 
  We see comparing with the results of the previous 
subsection, 
that the behaviour of the system for 
{\it strong} coupling is rather different from 
its behaviour in the weak coupling regime. For appropriate 
cavity dimensions, which are of the same order  of those ensuring stability in the weak 
coupling regime,
we ensure for strong coupling the complete decay of the particle to the ground state in a 
small ellapsed time. 

We have presented in this note a non-perturbative treatment of a 
confined {\it ohmic} quantum system consisting 
of particle (in the larger sense of a "material body", an atom or a Brownian particle) 
coupled to an environment modeled by non-interacting oscillators. 
One possible conclusion is that by changing conveniently the physical and geometric 
parameters (the emission frequency, the strenghness of the coupling and the size of the 
confining cavity) our formalism theoretically allows a control 
on the rate of emission and of  the energy storage capacity (perhaps information?) in the cavity. Depending on the strenghness of the coupling,
the emission probability ranges from a complete stability to a very
rapid decay. \\
 This work has been supported by the Brazilian agency CNPq (Brazilian 
National Research Council)


\begin{thebibliography}{99}


 
 
 
 
 
\bibitem{adolfo1} N.P.Andion, A.P.C. Malbouisson and A. Mattos Neto, J.Phys.{\bf A34},
3735, (2001)

\bibitem{adolfo2} G. Flores-Hidalgo, A.P.C. Malbouisson, Y.W. Milla,  Phys. Rev. A, {\bf 65}, 063314 (2002)

\bibitem{Haroche} S. Haroche, Doctoral thesis, Ecole Normale Sup\'erieure, Paris (1964) 

\bibitem{paz} B.L. Hu, Juan Pablo Paz, Yuhong Zhang, Phys, Rev. D, {\bf 45}, 2843 (1992)


\bibitem{Thirring}  W. Thirring, F. Schwabl, Er. Ex. Naturw. {\bf 36}, 219 (1964) 

\bibitem{Jhe} W. Jhe, K. Jang, Phys. Rev. {\bf A53}(2), 1126 (1996) 

\end{thebibliography}
\end{document}